\documentclass[12pt,preprint]{aastex}

\shorttitle{Gravitas}
\shortauthors{Dubinski \& Farah}

\begin{document}


\title{GRAVITAS: Portraits of a Universe in Motion}


\author{John Dubinski\altaffilmark{1}}
\affil{Department of Astronomy \& Astrophysics, 50 St. George St., 
University of Toronto, Toronto, ON, Canada M5S 3H4}
\email{dubinski@astro.utoronto.ca}

\and

\author{John Kameel Farah\altaffilmark{2}}
\affil{1147A Davenport Rd, Toronto, ON, Canada M6H 2G4}
\email{johnkameel@yahoo.com}


\altaffiltext{1}{website: http://www.galaxydynamics.org}
\altaffiltext{2}{website: http://www.johnkameel.com}


\begin{abstract}
GRAVITAS is a self-published DVD that presents a visual and musical celebration 
of the beauty in a dynamic universe
driven by gravity. Animations from supercomputer simulations of forming galaxies,
star clusters, galaxy clusters, and galaxy interactions are presented as moving
portraits of cosmic evolution. Billions of years of complex gravitational
choreography are presented in 9 animations - each one interpreted with an original
musical composition inspired by the exquisite movements of gravity. The result is
an emotive and spiritually uplifting synthesis of science and art.

The GRAVITAS DVD has been out for two years now but I am now making the DVD disk
image freely available for personal and educational use through a bittorrent 
download.  Download and burn at your leisure.  The animations are also downloadable 
in various video formats.  Follow the various links through the home website 
http://www.galaxydynamics.org/gravitas.html

\end{abstract}

{\em The liner notes of the DVD and track descriptions follow.}

\section{Understanding the Dynamic Universe}

When we look with wonder at galaxies in the beautiful images from space, we
immediately perceive that these vast systems of stars are in motion. The
spiral patterns of galaxies intuitively suggest a system in rotation. We
have all seen the spiral vortices of water going down the drain or similar
patterns in cream stirred into a cup of coffee. The distorted forms of
close pairs of galaxies with their long tails and interconnecting bridges
of stars are not as intuitive but do suggest that these systems are
interacting in a strong and violent way. We now know that these complex
patterns are manifestations of the long reach of gravity generated by the
stars, gas and dark matter that make up the galaxies. Unfortunately, we
cannot witness the dynamics of galaxies directly since the rotational times
are typically hundreds of millions of years, much longer than our
hundred-year life spans. Galaxies therefore appear as still-lifes to us -
incredibly beautiful images - but frozen snapshots presenting single frames
from a series of prolonged cosmic dramas.

The 9 computer animations presented on this DVD bring life to the snapshots
of galaxies and present the universe as a place that is in constant motion,
a place that is evolving and changing in complex ways over billions of
years. Each animation is a celebration of gravity acting in its purest form
through Newton's laws of motion. While Einstein usually gets most of the
credit for linking gravity to the larger universe, Newtons laws still reign
supreme for describing most of the behaviour in the solar system, star
clusters, galaxies and even bigger things like galaxy clusters and the
large-scale structure. Small, high-density objects such as neutron stars
and blackholes and the biggest thing we know, the universe seen as a whole,
require Einstein's detailed formulation of gravity in his Theory of General
Relativity. The phenomena of everything in between these two length scales
can be described very well by the methods of good old fashioned Newtonian
gravity.

The study of the motion of stars in clusters, galaxies and the large-scale
universe is known as stellar or galactic dynamics. The underlying
description is Newton's N-body problem: the computation of the motion of
"N" (a large number) bodies (e.g. planets or stars) each of which is
pulling and being pulled by every other body through the force of gravity.
The force between bodies is inversely proportional to square of the
distance between them. Newton's great triumph was the discovery of an exact
solution of the two-body problem for the inverse-square force law that
predicted that planets should move on Keplerian elliptical orbits. However,
for N greater than two the problem had no general solution. In the
pre-computer era, this problem was virtually intractable because of the
large amount of computing needed even for small systems of bodies. Only
complex perturbation calculations could be done to understand the
subtleties of the deviation from elliptical orbits of the inner planets due
to the gravitational effects of Jupiter and Saturn.

For a long time, the solar system was the only part of the universe that
was seen as dynamic and studied in detail with Newtons physics. However, as
telescopic observations expanded our view of the universe it became clear
that larger systems of gravitating bodies also occured in nature. Star
clusters containing hundreds and even as many as a million stars in
globular clusters were commonly observed. And finally astronomers realized
that the great spiral nebulae were actually large systems of stars often
laid out in a disk like the Milky Way but at such great distances that the
combined light of their stars appeared as a nebulous haze. These systems
contained hundreds of billions of stars and were configured as disks - much
like the layout of the solar system - and were spinning slowly in space.

A typical galaxy is one hundred thousand light years across with a mass of
one hundred billion suns implying rotational periods of hundreds of
millions of years - timescales so long that their spinning motions could
never be perceived in a human life span. Computing the motion of the 9
planets in the solar system was hard enough. How could one possibly deal
with one hundred billion stars?

In the pre-dawn light of the computer era, the Swedish astronomer Erik
Holmberg performed a remarkable experiment to explore the gravitational
dynamics of interacting galaxies. On a table top, he laid out two disk
configurations of 37 light bulbs, each one representing a spiral galaxy.
Light acted as a proxy for mass. Since the intensity of light falls off as
the inverse square of distance, the combined intensity of light measured by
a photocell placed at the position of a light bulb would be proportional to
the equivalent force of gravity. In this way, he meticulously measured the
"force" at the position of each light bulb which allowed him to calculate
the change in position according to Newton's laws and so solve the N-body
problem. By carefully moving the light bulbs according to this calculation
in a series of steps, he was able to follow the motion of this 74 particle
system. He showed that interacting galaxies could excite spiral structures.
The first studies in computational galactic dynamics had begun.

By the sixties, electronic computers became widely available to scientists
and early pioneers most notably Sverre Aarseth made a first hard attack at
numerical computation of the N-body problem modeling systems with a few
hundred particles. By the early seventies, the Brothers Toomre presented a
detailed collection of N-body calculations with accompanying elegantly
drafted drawings clearly illustrating that gravity was responsible for the
phenomenology of the tails and bridges in real interacting systems of
galaxies. As computers grew in power through the seventies and onward to
the end of the century, a variety of new efficient methods were developed
for computing the N-body problem to study the detailed dynamics of galaxies
and the growth of structure in the universe. The beautiful animation of the
merger of a small group of spiral galaxies created by Joshua Barnes in the
early nineties using the tree algorithm invented with Piet Hut led the way
inspiring further intense efforts to visualize galactic dynamics.

As we begin the 21st century, the methods for computing the N-body problem
have grown greatly in computational efficiency and have been adapted to
work on parallel supercomputers: vast arrays of thousands of computational
nodes interconnected with a high speed network. Calculations using tens to
hundreds of millions of particles are now routine and the animations
presented here are derived from simulations of this size. There are now
some calculations being done with as many as 10 billion particles! In a few
short years, it will be possible to compute the N-body problem in a
Galactic model containing as many particles as there are stars! This is a
truly amazing advance when looking back at Holmberg's heroic effort with 74
light bulbs.

The animations that I present here descend directly from this legacy of
hard work and ingenuity fueled by the desperate need to understand and see
the working machinery of nature at the largest scale. They build upon the
astrophysical knowledge and technical skills of many researchers over the
past few decades. Astronomers are driven to desperate acts by their
insatiable curiosity and voyeuristic tendencies. They want to know and they
like to watch. I hope you do too!

\section{Track Descriptions}

\subsection{Cosmic Cruise (1:55)}

About 14 billion years ago, the universe began in a Big Bang. In one single
instant, all matter and energy were created. Rapid expansion caused the
matter to cool and change into atoms and also the mysterious dark matter.
At first, the dark matter was spread out evenly but faint echoes of the
seething quantum foam that existed at the instant of creation remained like
random ripples on the surface of a frozen pond. Gravity took hold of these
noisy echoes and caused them to collapse into halos of dark matter that
became the seeds of the galaxies.

In this animation, we fly straight through a 130 million particle
simulation of dark matter travelling hundreds of millions light years over
14 billion years. We illuminate the dark matter particles so that we can
watch the formation of the cosmic web - the foundation of all structure in
the prevailing model of cosmology. At the start, the regular grid of
particles reflects the featureless nature of the universe at the beginning.
As the flight continues, we witness the formation of the first structures
through the collapse of density fluctuations. These merge with other
structures and grow into the dark halos of sizes varying from galaxies to
galaxy clusters.

\subsection{Galactic Encounters (3:11)}

The dark matter provides the framework for the universe but what we see are
the galaxies - vast islands of stars and gas that form at the centre of the
dark halos. The galaxies themselves can gather into enormous clusters with
hundreds and even thousands of members. There is little breathing room for
a galaxy in a cluster and soon strong interactions and collisions ensue as
the galaxies fall together. Galaxies are diaphanous objects - puffs of
smoke easily torn apart by the forces of gravity and many merge together
into an amorphous central blob of stars while others are left severely
damaged.

Here we watch a hundred galaxies fall together into a forming cluster. Our
perspective is from a starship flying into the cluster starting several
million light years away and cruising to within a hundred thousand light
years of the giant elliptical galaxy forming at the cluster centre. As we
fly through, we observe the merging and tidal disruption of many spiral
galaxies as they orbit within the cluster. Ten billion years elapses within
about 3 minutes so time passes at a rate of 50 million years per second!

\subsection{Swarm (2:14)}

By increasing the playback speed, we get a different impression of the
dynamical processes in clusters. Here we look at the cluster from the
previous track again but from 3 different fixed perspectives and also speed
up the playback to 200 million years per second. The cluster dynamics take
on a violent and frenzied character in stark contrast to our usual
impressions of stately slow changes from astronomical imagery.

\subsection{Nightfall (5:55)}

Not so long after the Big Bang, somewhere in the universe the first star
was born. Great clouds of gas condensing within the galaxies were stellar
cradles. The largest clouds condensed into globules that flared into
millions of stars and became globular clusters. These ancient families are
fossils of the first moments of creation and shine mainly today by the
light of red and blue giants. For billions of years, the stars have moved
on their way rarely encountering their companions shuttling back and forth
indifferently guided by the relentless force of gravity.

This animation is inspired by the classic Isaac Asimov short story
Nightfall written in Astounding Stories in 1941. There an astronomer on a
planet with multiple suns where it is always day learns through his N-body
calculations that all of the suns will soon set for the first time in
thousands of years. The last time this happened civilization collapsed
because of the mass insanity that followed when night fell. As the last sun
sets and mass hysteria ensues, the astronomer looks up into the sky and
perceives the fantastic view of a sky filled with tens of thousands of
stars for he lives on a planet at the heart of a globular cluster! I have
visualized the stars in the animation to reflect the true colour and
brightness distribution in globular clusters and reproduce the look of
Hubble Space Telescope images of globular star clusters accurately.

\subsection{Klemperers Dream (4:03)}

Nature provides us with random acts of gravitational violence in the form
of galaxy collisions. Graceful spiral forms, tails and bridges are sculpted
by these interactions. But we know how gravity works and galaxies are
structured, so we can take the sculptors tool from Natures hand and play
with it in a supercomputer removing the random element.

Klemperer found that special symmetric arrangements of particles could
follow predictable orbits. These exact N-body solutions seem to have no
natural counterpart and even Klemperer stated that he really just studied
them for fun! In the same spirit, I have put galaxies in similar unnatural
symmetric configurations to explore their evolution. One amazing
consequence of Newtons laws of motion is that any system with some symmetry
built in should preserve that symmetry even if complex dynamical behaviour
is occuring. For the sequence of simulations here, it seems that symmetry
is preserved even when spiral patterns emerge after the galaxies interact
strongly. But by the end of each sequence, the symmetry is lost. So is
Newton wrong afterall? No! These nonlinear dynamical systems are unstable
and become chaotic. Tiny deviations introduced by computer imprecision are
eventually amplified and lead the system away from symmetry. So these
simulations are not just for fun after all. They are an interesting
illustration of the emergence of chaos in nonlinear dynamical systems.

\subsection{Spiral Metamorphosis (6:18) /  Future Sky (6:37)}

The harsh reality of the distant universe with all of its violent
interactions seems remote from our human existence and all might seem to be
quiet and normal in our home the Milky Way. But it seems likely that in a
mere 3 billion years, our neighbouring galaxy Andromeda and the Milky Way
will fall together and have a close collision. They will likely merge and
be reborn as a single giant elliptical galaxy over the course of another
billion years or so. How might this metamorphosis play out and what might
you see if you looked up at night over the next 4 billion years! The space
between stars is so vast compared to their size that during a galaxy
collision no individual stars actually collide with one another. So our sun
and its family of planets will be taking a passive but exciting ride
through the pair of coalescing galaxies and take on a spectacular view of
the unfolding disaster in relative safety.

I have set up a model system of colliding galaxies that reflects the
current state of our the Milky Way and Andromeda system. There are still
some uncertainties about the exact trajectories and masses of the two
galaxies but I have set up a plausible case where they fall together and
collide almost directly passing within 60000 light years of each other.
Also, I only present the view of the naked stars unobscured by the
interstellar gas and dust clouds within the galaxy.

We get a chance to see it all from 4 perspectives: two fixed positions in
space a million light years away (Spiral Metamorphosis) and two sky views,
the first that projects the full 360 degrees of the sky onto an oval map
and the second a view of one hemispheric dome of the night sky. In the sky
views (Future Sky), one particle is identified as the sun within the model
of the Milky Way and our view is always from this perpective with our
attention directed towards the central bulge of the Galaxy making for a
mind boggling spectacle.

The view from far reveals an exquisite ballet of mutual annihilation and
transformation into an elliptical galaxy. The Milky Way is seen coming in
from the bottom in a face-on and edge-on view. After the interaction, long
tidal tails of stars are flung out in open spiral patterns from both
galaxies by the strong gravitational tides during the interaction. While
separating, the two galaxies develop detailed spiral structure and then
fall back for a second collision finally to merge. The mutual annihilation
of the two galaxies leads to a big splash showing up as a complicated
system of loops and ripples that represent turning points of stellar
orbits. The two galaxies finally settle down into a single elliptical
galaxy surrounded by remnant debris of their violent interaction.

In the sky views, the arch of the Milky Way is apparent at first as a band
of stars and tiny Andromeda is seen scrolling past beneath the arch but
slowly growing in size as it approaches. When the 2 galaxies intersect, the
sun is flung out far from the colliding pair of galaxies and our view
oscillates between a remote view of events to a wild ride right through the
centre of the galactic bulge! The orbit of the sun is no longer circular
but now follows a convoluted pattern with the distorted gravitational field
of the merging galaxies. A final look back from the far flung sun shows the
final merger of the two galaxies.

\subsection{Galaxies in Collision (7:53)}

We revisit the galaxy cluster simulations of Swarm followed by random views
of two-galaxy collisions. The animation begins slowly to allow an
appreciation of the grace of gravitational dynamics but the tempo gradually
increases to change the impression to a frenetic dance of severe intensity
and violence. The animation ends with a fireworks display of rapid fire
merging pairs of galaxies fading into the Hubble Space Telescope image of
the The Mice one of the well-known nearby famous pairs of interacting
galaxies. The final image reminds us that we live in one single instant of
an evolving universe.

\subsection{Metamorphosis 3-D (6:13)}

I have re-run the simulation of Spiral Metamorphosis using stereoscopic
rendering methods to produce a red-blue 3-D animation. To appreciate this
animation, put on the 3-D glasses (red lens over your left eye) and seat
yourself in direct line with the centre of the screen at a distance of
about 5 times the width of your screen. Relax your eyes and watch a
3-dimensional rendition of the merger of the Milky Way and Andromeda from 3
different perspectives! In 3-D, the patterns of shells and ripples take on
a completely different light and reveal the amazing depths of complexity in
the dynamics of galaxies.

\section{About the Music}

An essential component of this compilation of animations is the music of
composer-pianist John Kameel Farah. He draws upon and synthesizes the
sound-worlds of renaissance and baroque counterpoint, free improvisation,
Middle-Eastern music, ambient minimalism, techno and electronica to create
a music that crosses time and dimension. His creative efforts are fueled by
exchanges of physical, spiritual and emotional energies, on both a
macro-cosmic and microscopic level. The accompanying compositions are
direct musical interpretations of the unfolding dynamics presented in each
animation. The music provides a wondrous ambience and serene state of mind
that permits us to contemplate the beauty of a universe in motion. For more
information on concerts, recordings and projects, please visit
www.johnfarah.com.

\acknowledgments

Financial support from Science and Engineering Research Canada, the
Canadian Foundation for Innovation, the Canadian Institute for Theoretical
Astrophysics and the University of Toronto is gratefully acknowledged. JD
would like to thank Frances, Alicia and Kintaro for their love, support and
patience over the years and he would not have started down this road long
ago without the backing of his loving parents Margaret and Roman. JD also
tips his hat to James Binney and Scott Tremaine for teaching him much of
what he knows about galactic dynamics from their book and Josh Barnes and
Piet Hut for inventing the N-body tree algorithm. JKF would like to thank
his parents Nadia and Farid Farah, and A.G. Swayze for support and
inspiration.

%\begin{figure}
%\epsscale{0.8}
%\plotone{cover.eps}
%\caption{Gravitas DVD Cover}
%\end{figure}

%\begin{figure}
%\epsscale{.68}
%\plotone{poster.eps}
%\caption{In 3 billion years, the Milky Way and Andromeda galaxies
%will likely collide.  The two galaxies are shown here a few hundred million years
%after closest approach with the smaller Milky Way on the bottom.  The strong
%gravitational tides of the interaction excite open spiral arms that develop into
%sweeping tidal tails and bridges.  Eventually, the two galaxies will merge and
%transform in a single elliptical galaxy.  This simulation used more than 300
%million particles to represent the stars and dark matter in the two galaxies and
%was calculated with 512 processors on the parallel supercomputer Mckenzie at the
%Canadian Institute for Theoretical Astrophysics in Toronto in 2003.  This poster
%is can be downloaded at full resolution at
%http://www.galaxydynamics.org/dvd/poster.html }
%\end{figure}

\end{document}